# Dual-domain Microwave Photonic Radar for Non-contact High-resolution Monitoring of Vital Signs in Multiple Individuals

Dongyu Li[†], Ziqun Zhang [†], Hongyi Wang, Yukang Jia, Siyue Zeng, Hong Chen, Jin Zhang, Xiaotong Liu, Yalan Wang, Dangwei Wang, and Anle Wang[*]

*Early Warning Academy, Wuhan 430019, China*
[†]*These authors contributed equally.*
[*]*anlehit@163.com*

**Abstract:** We present a dual-domain microwave photonic radar system for multi-parameter vital sign monitoring, utilizing ultra-wideband radiofrequency signals to achieve high-precision chest displacement measurements, while high-frequency optical signals enable accurate Doppler detection of pulse-induced micro-motions. We experimentally validated the system by simultaneously monitoring respiratory and pulse rate of two male volunteers, demonstrating high performance with average accuracies of 98.1% for pulse detection and 99.85% for respiratory detection. Additionally, the system facilitates rapid, indirect blood pressure estimation, achieving a coefficient of determination ($R^2$) of 0.905. With the help of low-loss distribution capability of optical fibers, the system can be scaled into a distributed, non-contact multiple vital signs monitoring platform, highlighting its potential for clinical and healthcare applications.

## 1. Introduction

Vital signs, including respiratory rate, pulse rate, body temperature, and blood pressure, serve as fundamental indicators of human physiological functions [1]. Continuous monitoring of these parameters is essential for real-time assessment of an individual's health conditions. With the global population aging rapidly, round-the-clock vital signs surveillance has become particularly critical for elderly care, enabling prompt medical intervention when abnormalities are detected by immediately alerting healthcare providers or family members. In addition, continuous non-contact multiple vital signs monitoring contributes significantly to public safety maintenance by facilitating early risk identification. Research indicates that individuals with malicious intent often demonstrate characteristic physiological responses, such as tachycardia or hypertension, prior to committing harmful acts [2]. The implementation of non-contact vital sign detection technology allows for proactive identification and continuous behavioral monitoring of potential threats, thereby enhancing preventive measures and minimizing possible harm to public safety.

Current non-contact vital sign detection primarily relies on two approaches: imaging-based and radar-based systems. Imaging-based systems raise privacy concerns and are susceptible to ambient lighting conditions, which restrict their application scenarios [3,4]. Radar-based systems can detect vital signs remotely using radiofrequency (RF) waves rather than camera recording, ensuring both non-contact operation and user privacy protection [5]. Existing electronic radar systems can extract vital sign information, such as respiration and heartbeat [6-11]. Conventional electronic radar systems are limited by their RF signal bandwidth [12,13], resulting in insufficient range resolution to directly detect tiny chest displacement of ~1 cm caused by human respiration [14]. Current approaches typically rely on complex signal processing algorithms to extract respiratory rate and heart rate information from the echo signal. Besides, most existing frequency-modulated continuous-wave (FMCW) radar-based heartbeat detection systems exhibit significant limitations: the acquired cardiac signals lack clearly defined fiducial points marking the onset and termination of individual heartbeats, permitting only time-averaged heart rate estimation [15-17]. Although Wang et al employed variational mode decomposition (VMD) to extract higher-quality heartbeat signals from micro-Doppler signatures and achieved inter-beat interval (IBI) estimation through envelope peak detection [18], their method still fails to capture clinically relevant cardiac cycle features such as distinct systolic and diastolic phases or blood pressure. Microwave photonic (MWP) radar technology presents a solution by overcoming conventional bandwidth limitations and achieving millimeter-level resolution [19]. In 2023, Zhang et al. proposed a microwave photonic radar system [20], which generated a 10 GHz bandwidth stepped-frequency RF signal in the Ka band, achieving a distance resolution of 13.7 mm and millisecond-level measurement accuracy of the respiratory rate of cane toads. However, their study was confined to single-target, non-human vital sign detection, leaving broader biomedical applications unexplored.

In this study, we present a dual-domain microwave photonic radar system capable of non-contact, simultaneous monitoring of respiratory and pulse rates of multiple individuals. The system utilizes an optical frequency operation module (OFOM) [21] to generate ultra-wideband RF signals, enabling precise detection of chest wall displacements associated with respiration. Meanwhile, by extending an optical branch based on the same OFOM and leveraging the high-frequency characteristics of optical signals, the system achieves accurate extraction of Doppler signatures corresponding to subtle pulse movements. The proposed system combines the complementary features of microwave and optics, and the vital sign information detected by these two modalities can be cross-verified, which can improve the measurement accuracy and provide individual health information. We evaluated the performance of the proposed system through comprehensive experiments involving both simulated vital signs and human vital signs. The experimental results demonstrated that the system has high performance in measuring respiratory rate and pulse rate. Furthermore, our findings suggest that the proposed method can provide rapid, indirect assessment of blood pressure levels, with a coefficient of determination ($R^2$) of 0.905. Fully exploiting the ubiquitous distribution of microwaves, the low-loss distribution characteristics of optical fibers, and the advantage of the de-chirping receiving technology that can convert broadband high-frequency signals into low-frequency single-frequency signals, a distributed and non-contact multiple vital signs detection system is also proposed. The ultra-wideband transmitted signal enables the system to possess a high resolution of frequency, allowing precise discrimination of multiple individuals' vital signs in the frequency domain. To demonstrate the practical application, the distributed system successfully accomplished simultaneous measurement of respiratory and pulse rates for two volunteers, with average measurement accuracies of 99.85% and 98.1% respectively, demonstrating its multi-individual monitoring capability. The proposed distributed system can be applicable to various scenarios such as hospitals, nursing homes, and home monitoring, demonstrating broad application prospects. This research establishes the connection between microwave photonics and biomedical photonics, revealing the significant potential of microwave photonic radar in advancing biomedical photonics and non-contact medical diagnostics.

## 2. Principle

Fig. 1(a) shows the schematic of the proposed dual-domain MWP radar system for multiple vital signs measurement. The system is mainly composed of the OFOM, the optical signal transmission and reception module (OTRM), and the microwave signal transmission and reception module (MTRM). The OFOM generates an optical signal $E_{\text{out}}(t)$ containing two symmetric chirped sidebands, which is subsequently split into the OTRM and MTRM, respectively. In the OTRM, the optical signal is radiated via a collimator to detect the pulse. The collimator acts as a sensing access point, which transmit and receive the optical signals in the same channel. While in the MTRM, the optical signal first converts into electrical signal through a photodetector (PD), followed by free-space radiation using an antenna for respiratory monitoring. The transmission and reception of RF signals are implemented by using separate antennas, effectively avoiding electromagnetic interference between signals.

The optical signal generated by the OFOM can be expressed as

$$E_{out}(t) = E_1 \left[ \exp\left(j2\pi f_{c-}t + j\pi k t^2\right) + \exp\left(j2\pi f_{c+}t - j\pi k t^2\right) \right] \tag{1}$$

where $E_1$ represents the amplitude of two optical sidebands, $f_{c-} = f_c - f_{RF}/2$, $f_{c+} = f_c + f_{RF}/2$, $f_c$ represents the frequency of the laser source in OFOM, $f_{RX}$ represents the frequency difference between two optical sidebands, and $k$ represents the chirp rate of the optical sidebands.

For the OTRM, the optical signal output from the collimator can be expressed as

$$E_T(t) = E_2 \left\{ \exp\left[ j2\pi f_{c-}(t-\tau_1) - j\pi k(t-\tau_1)^2 \right] + \exp\left[ j2\pi f_{c+}(t-\tau_1) + j\pi k(t-\tau_1)^2 \right] \right\} \tag{2}$$

where $E_2$ represents the amplitude of the optical signal, $\tau_1$ represents the link delay from the OFOM to the collimator. Supposing that the link delay from the OFOM to the OC is $\tau_2$ and disregarding the link attenuation, the local optical signal transmitted to the OC from the OFOM can be expressed as

$$E_L(t) = E_3 \exp\left\{ \left[ j2\pi f_{c-}(t-\tau_2) - j\pi k(t-\tau_2)^2 \right] + \left[ j2\pi f_{c+}(t-\tau_2) + j\pi k(t-\tau_2)^2 \right] \right\} \tag{3}$$

The human pulse movement is caused by the contraction and dilation of blood vessels due to the heartbeats [22]. Consequently, there will be periodic displacement alterations in the skin adjacent to the artery. When the output optical signal of the collimator irradiates the skin near the artery, the echo optical signal can be expressed as

$$E_R(t) = E_4 \left\{ \begin{array}{l} \exp\left[ j2\pi(f_{c-} + f_d(t))(t - \tau_1 - \tau(t)) - j\pi k(t - \tau_1 - \tau(t))^2 + \phi \right] \\ + \exp\left[ j2\pi(f_{c+} + f_d(t))(t - \tau_1 - \tau(t)) + j\pi k(t - \tau_1 - \tau(t))^2 + \phi \right] \end{array} \right\} \tag{4}$$

where $E_4$ is the amplitude of the echo optical signal, and $\phi$ is the phase variance caused by the skin reflection. $\tau(t)$ is the time delay caused by the periodic displacement of the pulse, which can be expressed as

$$\tau(t) = \frac{2d_0}{c} + \frac{2d(t)}{c} \tag{5}$$

Here, $d_0$ represents the distance between the skin and the collimator assuming that the skin is static, $d(t)$ denotes the time-varying displacement caused by tiny vibration of skin, $c$ is the speed of light. The range resolution is defined as the minimum distance between different objects FMCW radar can distinguish, given by $r_c = c/2B$, where $B$ is the bandwidth of the optical signal. For human pulse detection, the skin displacement caused by pulse is less than 1 mm. Since the tiny vibration $d(t)$ do not cause skin to move beyond its range resolution, the second term in Eq. (5) can be neglected. $f_d(t)$ is the Doppler shift caused by tiny vibration of skin, which can be expressed as

$$f_d = \frac{2v(t)f}{c} \tag{6}$$

Here, $v(t)$ represents the radial velocity of human skin movement induced by vital signs, and $f$ denotes the frequency of the signal used for detection. Pulse detection in this system utilizes optical signals with the frequency of approximately 193.5 THz. Compared to microwave signals, which operate at frequencies on the order of ~GHz, the significantly higher frequency of the optical signal effectively enhances the Doppler shift, improving detection sensitivity.

The echo optical signal and the local optical signal are coupled by the OC and then frequency-mixed by the PD1 to achieve the de-chirp processing. The intermediate frequency (IF) output signal from PD1 can be represented as

$$I_L(t) = A_L \cos\{2\pi\{k[\tau_1 - \tau(t) - \tau_2] - f_d(t)\}t + \phi_1\} + A_L \cos\{2\pi\{k[\tau_1 + \tau(t) - \tau_2] + f_d(t)\}t + \phi_2\} \quad (7)$$

where $A_L$ the amplitude of the intermediate frequency signal after de-chirping, $\phi_1$ and $\phi_2$ are the phases of the echoes. Then, the IF signal frequencies of OTRM are respectively expressed as

$$\begin{aligned} f_{L-}(t) &= k[\tau_1 - \tau_2 - \tau(t)] - f_d(t) \\ f_{L+}(t) &= k[\tau_1 - \tau_2 + \tau(t)] + f_d(t) \end{aligned} \quad (8)$$

Let $\tau = \tau_1 - \tau_2 + 2d_0/c$, and the Eq. (8) can be simplified as

$$\begin{aligned} f_{L-}(t) &= k\tau - f_d(t) \\ f_{L+}(t) &= k\tau + f_d(t) \end{aligned} \quad (9)$$

For MTRM, the double-sideband broadband optical signal output by OFOM is converted into an ultra-wideband RF signal by the PD2, and then amplified by a power amplifier and propagated outward via the antenna. The RF signal can be expressed as

$$E_{RF}(t) = E_5 \exp\left[j2\pi f_{RF}(t - \tau_1') - j\pi k(t - \tau_1')^2\right] \quad (10)$$

where $\tau_1'$ represents the delay of the electrical signal during link transmission.

Human breathing leads to periodic displacements of the thoracic cavity. During the respirate detection, the received RF echo signal can be expressed as

$$E_{RX}(t) = E_6 \exp\left[j2\pi(f_{RF} + f_{dR}(t))(t - \tau_1' - \tau_R(t)) + j\pi k(t - \tau_1' - \tau_R(t))^2 + \varphi\right] \quad (11)$$

where $E_6$ is the amplitude of the echo signal, $\tau_R(t)$ is the delay between transmitting and receiving RF signals, $f_{dR}(t)$ is the Doppler shift caused by displacement of the chest wall and $\varphi$ is the phase change caused by the reflection. The received RF signal is input to the MZM and the output signal of the MZM can be expressed as

$$E_M(t) = E_{out}(t - \tau_2') \cdot \cos\left(\frac{\pi}{8} + \frac{\pi E_{RX}(t)}{2V_\pi}\right) \quad (12)$$

where $\tau_2'$ is the delay of the output signal from the OFOM to the MZM through the link, $V_\pi$ is the half-wave voltage of the MZM. The MZM operates at the quad point, and only the $\pm 1$ order sidebands are retained in the output of the MZM. Then the frequency of the IF signal output from the PD3 can be expressed as

$$f_R(t) = 2k[\tau_1' - \tau_2' + \tau_R(t)] + f_{dR}(t) \quad (13)$$

The MTRM use microwave signal for respiration detection with a frequency in the GHz range, producing Doppler shifts at the Hz level, which is negligible compared to the first term in Eq. (13). $\tau_R(t)$ varies with time dues to the movement of the respiration, which can be expressed as

$$\tau_R(t) = \frac{2d_0'}{c} + \frac{2d'(t)}{c} \quad (14)$$

where $d_0'$ denotes the distance between the chest and the antenna in the absence of respiratory, and $d'(t)$ represents the time-varying displacement of the chest induced by respiration. The proposed system in this research employs ultra-wideband microwave signals for detection, achieving millimeter-level range resolution. This resolution is sufficient to resolve centimeter-scale chest displacement induced by respiration. Consequently, unlike the case of pulse detection, the second term in Eq. (14) cannot be neglected. Let $\tau' = \tau'_1 - \tau'_2 + 2d'_0/c$, and the Eq. (13) can be simplified as

$$f_R(t) = 2k\left[\tau' - \tau_R(t)\right] \tag{15}$$

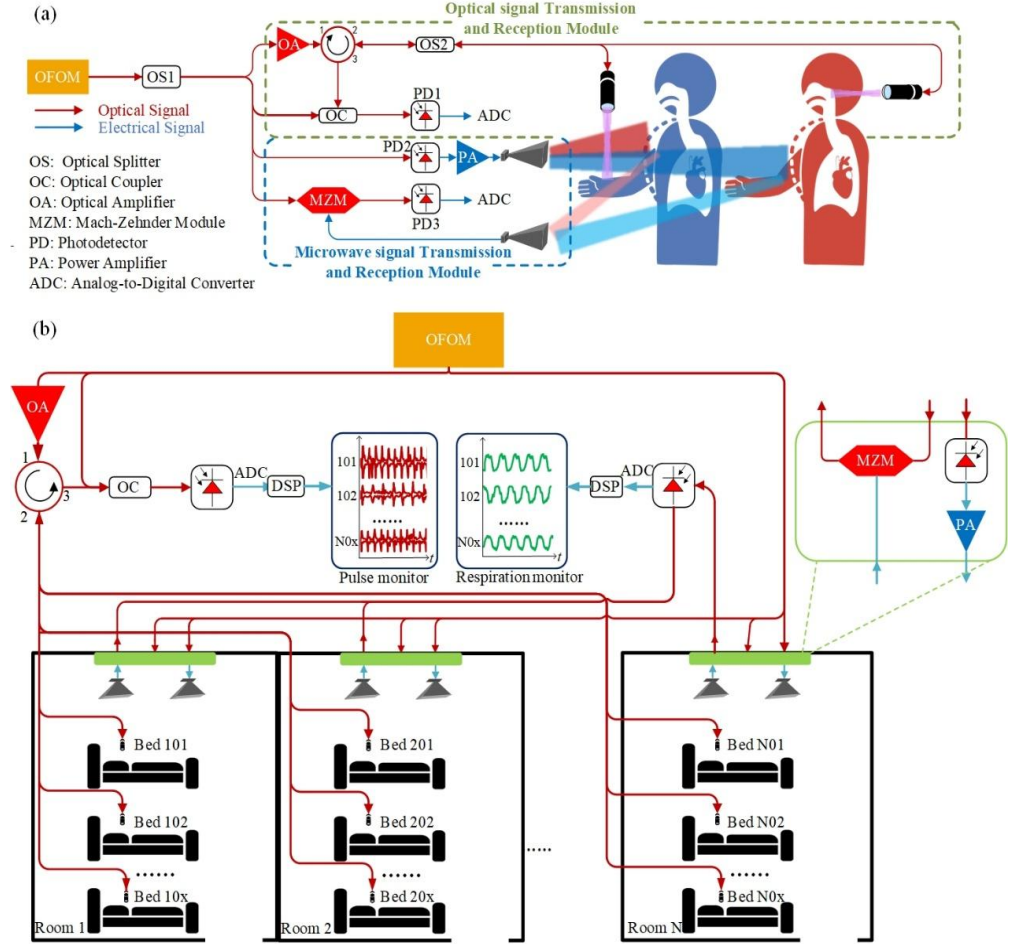

Fig. 1. The dual-domain MWP radar system for non-contact multiple vital signs detection. (a) The schematic of the dual-domain MWP radar system; (b) Conceptual drawing of the distributed vital sign radar system with distributed sensing access points. The blue lines represent RF coaxial cables, while the red lines represent optical fibers.

The Eq. (9) and (15) show that the de-chirped frequencies of the IF signals demonstrate periodicity that matches the rhythms of the vital sign. By extracting the time-varying frequency component from the echo IF signal, the periodic variations of vital signs can be characterized.

Eq. (9) and (15) also show that the center frequencies of IF signals change with the echo delay $\tau$ and $\tau'$. For the OTRM, changing the length of the optical fiber connected to the collimator can alter the echo delay, which further change the center frequency of the IF signal. Therefore, the system will have more access points to enable simultaneous measurement for multiple individuals by setting different lengths of the optical fiber connected to each collimator. For the MTRM, the distances between the objects and antenna are different, resulting in different echo delays. Thus, the MTRM naturally has the characteristic of simultaneously measuring the respiration of multiple individuals. Compared to the RF cables with the attenuation over 50 dB/km at GHz frequencies, optical fibers have superior transmission performance with the attenuation of 0.2 dB/km, which are particularly suitable for distributed transmission systems. A distributed and non-contact multiple vital signs detection radar system has been proposed through optimized fiber length design as shown in Fig.1 (b). By utilizing low-loss optical fibers to link multiple sensing access points (collimators and antennas) to each room, the system enables simultaneous monitoring of multiple vital signs across multiple rooms. The upper bound of individuals for simultaneous monitoring are fundamentally limited by

$$N = \min\left(\frac{B_0 T c}{4 B_1 n \Delta L}, \frac{B_0 T c}{8 B_2 \Delta d}\right) \quad (16)$$

where $B_0$ is the sampling rate of the analog-to-digital converters (ADC), $B_1$ is the bandwidth of the optical signal, $B_2$ is the bandwidth of the microwave signal, $n$ denotes the refractive index of the optical fiber (typically ~1.45 for standard single-mode fiber), $T$ is the signal period and $c$ is the speed of light. This system is applicable to various scenarios such as hospitals, nursing homes, and home monitoring, demonstrating broad application prospects.

## 3. Materials and methods

An experimental system is constructed to verify the effectiveness of the dual-domain MWP radar system, according to the Fig.1(a). The proposed system was constructed based on the OFOM module in [21], while the only difference is that the dual-output RF signal source in the OFOM is replaced by a customized product (PSS5R8C). Key devices and relative key parameters used in the experimental system are shown in Table 1.

Table 1 Information and key parameters of the main components of the experimental system

| Instrument Name | Model and Manufacturer | Key parameters |
|---|---|---|
| dual-output RF signal source | Sevenstar. PSS5R8C | CH1: Frequency 39.5 GHz<br>CH2：LFM, Center frequency 18 GHz，Bandwidth 1.25 GHz，Repetition period 300 μs. |
| PD1&PD3 | Discoveryoptica, PDA2GC | Responsivity: 0.9 A/W<br>3dB Bandwidth: 2GHz |
| PD2 | Finisar, XPDV2120RA | Responsivity: 0.69 A/W<br>3dB Bandwidth: 50 GHz |
| MZM | EOSAPCE. AX-0MSS-40 | 3dB Bandwidth: 40 GHz |
| Power amplifier | HONGYU CHENGDIAN-STAR | Frequency: 18-40 GHz<br>Gain ≥40dBm<br>saturated output≥37.5 dBm |
| Circulator | OVLink. COL-1550 | Operating wavelength:1550nm |
| Electrically controlled track | Daheng Optics, GCD-501100M | Resolution: 0.5 μm<br>Position accuracy: 10 μm |

The IF signals are digitized by the ADC and subsequently filtered through a bandpass filter to filter out the specific IF signals of each channel. The short-time Fourier transform is then applied to obtain the time-dependent frequency variations of the IF signals within each channel. As derived from Eq. (9) and (15), these variations of frequency correspond to periodic physiological movements associated with pulse and respiration. The pulse rate and respiratory rate are determined by quantifying the number of such periodic movements occurring within one minute.

In this work, the sampling rate of the ADC is 1 GHz, with optical and microwave signal bandwidths of 10 GHz and 20 GHz, respectively, and a signal period of 300 μs. The system theoretically supports simultaneous monitoring of at least 517 individuals for both respiratory and pulse rate.

## 4. Results

*4.1 Simulation verification experiment*

To verify the effectiveness of the system, we conducted a series of experiments. As shown in Fig. 2(a), the optical signal generated by the OFOM exhibits a signal-to-noise ratio (SNR) exceeding 40 dB, sideband bandwidth of which is up to 10 GHz. After the PD 2, the spectrogram of the generated K-Ka RF signal is illustrated in Fig.1(b). The RF signal is centered at 28 GHz with a bandwidth of 20 GHz, leading a theoretical range resolution of 7.5 mm.

The simulation verification experiment based on standard samples was first completed to test the system's perception ability for small displacements and micro-motions. A metal plate reflector (20 cm×20 cm) was installed on an electrically finely controlled track to simulate the movement caused by human breathing and pulse. The OTRM and MTRM were used to detect the minor displacement of the simulator, as shown in Fig. 2(c). The reflector was fastened on the motor and positioned 5 mm ahead of the collimator. Firstly, the motor was configured to drive the reflector to reciprocate within a 5 mm range, simulating the movement of pulse. The displacement of 5 mm is within the range resolution capability of OTRM, thus, what was measured is the periodic Doppler velocity variation induced by the periodic movement of the reflector. The motion period of the reflector was set at 2 seconds, with each of the forward and reverse movements lasting for 1 second, respectively. The maximum movement velocity was configured as 23.7 mm/s. The maximum Doppler frequency of the echo signal received by the OTRM was approximately 0.03 MHz, and the corresponding velocity of the reflector was 23.3 mm/s as calculated. Experimental observations also indicated that the IF echo signal presented significant periodic features, and its real frequency variation in one period could be characterized by the envelope indicated by the red line in the Fig. 2 (d), with a period of approximately 2 seconds. It is worth noting that the IF echo signal shows excellent symmetry, which is in good consistency with the theoretical derivation Eq. (9). This phenomenon stems from the transmitted signal characteristics adopted by the OTRM: the transmitted signal consists of two chirp signals with opposite frequency modulation slopes and the center frequencies of these two chirp signals are symmetric about the carrier frequency. Subsequently, the reflector was placed approximately 135 cm in front of the antenna. The motor drove the reflector to conduct periodic reciprocating motion, with a reciprocating distance of 2 cm and a period of 3.4 seconds. Since the reflector has a larger physical size rather than being an ideal point target, the beam emitted by the antenna is reflected not only by the reflector but also by the mechanical fixture structure that moves synchronously with the reflector. The echoes from these different reflective bodies are all captured by the receiving antenna, resulting in the IF signal composed of multiple frequency component variations rather than a single frequency curve. Notably, due to the identical motion trajectories of the reflector and the fixture, the variation patterns of each frequency component exhibit a high degree of consistency. As shown in Fig. 2(e), 8.5 cycles can be observed within the 30-seconds observation period, which is in accordance with the set parameters. These demonstrated results indeed prove that the proposed system can be utilized to measure micro-motion and micro-Doppler information, which make it possible to detect the respiration and pulse of human beings.

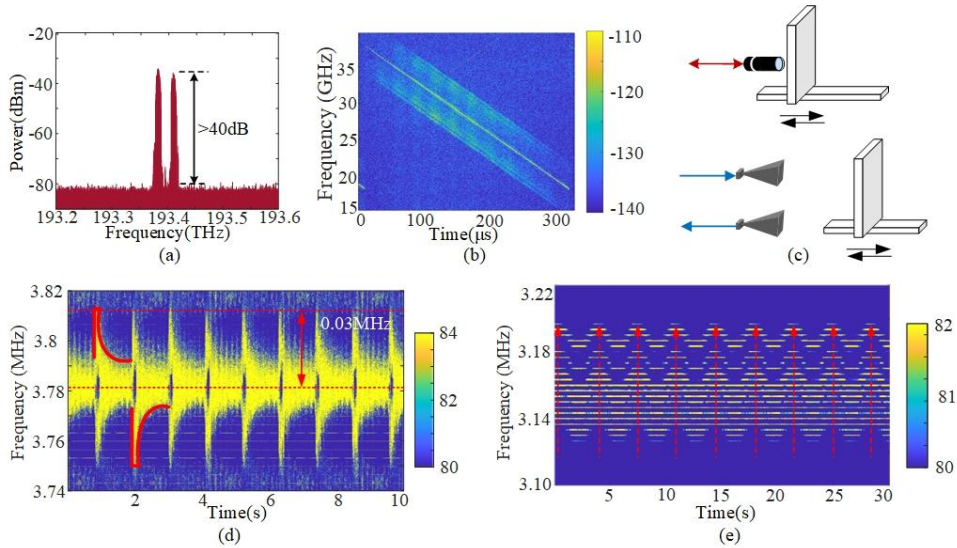

Fig. 2 The output signal of the dual-domain MWP radar system and the vital sign detection results based on simulators. (a) The optical spectrum of the output optical signal. (b) The STFT spectrogram of the output microwave signal. (c) The schematic of the vital sign detection based on simulators. (d) The STFT spectrogram of the IF echo signal received by the OTRM. The red line indicates the frequency variation in one period. The maximum Doppler frequency was approximately 0.03 MHz (e) The STFT spectrogram of the IF echo signal received by the MTRM. Each red dashed arrow represents one motion cycle.

*4.2 Human verification experiment*

To further evaluate the system performance under realistic conditions, we substituted the metal plate reflector with a 22-year-old healthy male volunteer for respiratory rate monitoring. The respiratory signal was first recorded while the volunteer was at rest as shown in Fig.3(a). Then, the volunteer performed an open-close jump exercise for 2 minutes. The respiratory signal was detected immediately after the exercise, as shown in Fig. 3(b). The received echo signal demonstrates periodic oscillations corresponding to chest wall movements induced by respiration. The de-chirped IF signals exhibit multiple frequency components, which arise from the non-ideal single reflection surface of the human thoracic wall that actually contains multiple scattering points. All frequency components demonstrate identical periodicity characteristics, from which we extract the respiratory rate using the component with the optimal SNR. As indicated by the red arrows in the figure, each marker represents one complete respiratory cycle, and the respiratory rate is determined by counting the number of cycles within 60 seconds. The respiratory rate of this volunteer increased significantly from 18 breaths per minute at rest to 35 breaths per minute after physical exercise. Local signal amplification and amplitude extraction demonstrated a marked rise in respiratory signal magnitude from 45 kHz to 121 kHz after exercise, indicating substantially enhanced thoracic movement. These results align with the expected physiological responses. Notably, subtle frequency shifts are observed in the median value of the IF signals, attributable to slight body movements of the subject during measurement. Furthermore, variations in frequency corresponding to the optimal SNR component are detected between pre- and post-exercise measurements, primarily caused by minor differences in the distance between the volunteer and the antenna. However, these factors do not compromise the accuracy of respiratory rate measurement.

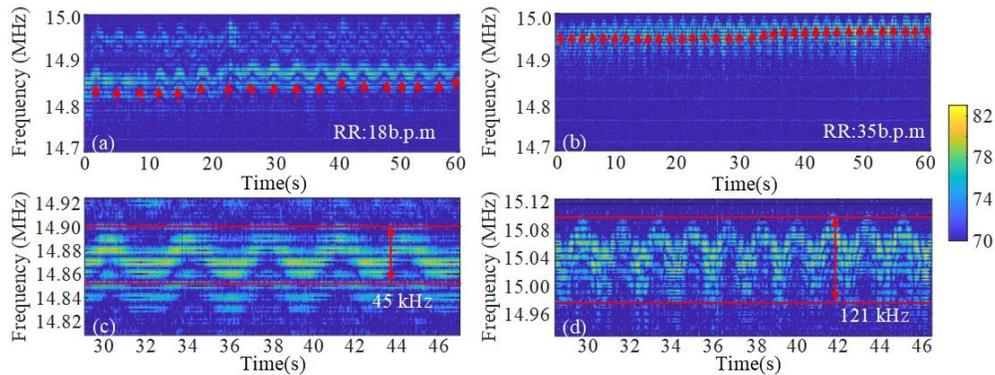

Fig. 3 Respiratory monitoring results of the same individual at rest (a) and after exercise (b). The white numbers indicate the measured respiratory rates (RR) in breaths per minute (b.p.m).

Existing studies have demonstrated that physiological processes such as blood circulation and muscle tremors can induce micro-motions in human epidermis [23,24]. However, the pulsation of superficial arteries generates more pronounced cutaneous vibrations. To verify whether the detected micro-motion signals of human epidermis originate from arterial pulsation, pulse measurements were performed at 17 predefined locations around wrist, following a grid pattern with 5 mm spatial resolution, as shown in Fig. 4(a). Continuous pulse monitoring was conducted under resting conditions, with positional adjustments made at 30-seconds intervals. The acquired pulse wave signals from these measurement points are presented in Fig. 4(b). Comparative analysis demonstrated significantly higher pulse amplitudes along line A compared to lines B and C, suggesting that line A closely aligns with the radial artery, while lines B and C are positioned at progressively greater distances from the radial artery, resulting in attenuated cutaneous vibrations. Notably, weak pulses could be detected in certain regions of line B, while the obvious pulse signals were hardly observable along line C. This spatial variation in signal intensity may stem from differences in subcutaneous tissue composition. The muscle tissue beneath line B is more responsive to pulse vibrations, whereas line C, being closer to the tibia, experiences less influence from the pulse vibrations. Clear pulse signals were detected at positions A1-A5, which correspond to the Cunkou positions described in traditional Chinese medicine (TCM), as depicted in Fig. 3(c). In contrast, positions A6 and A7 exhibit less pronounced pulse signals due to their greater distance from the radial artery. The Cunkou diagnostic method is one of the most common methods of pulse collection in the clinical diagnosis of TCM. Practitioners can determine the physiological and pathological state of the body by feeling the pulse beating at the Cunkou [25]. The experimental results indicate that the characteristic pulse waveforms at different positions around Cunkou can be identified, potentially offering novel methodologies for pulse acquisition and diagnostic evaluation in future research applications. The experimental results indicate that position A2 exhibited the strongest pulse vibration amplitude, with progressively decreasing intensity observed in surrounding locations. Consequently, all subsequent pulse measurements around wrist in this study were conducted at position A2 and similar position optimization has also been carried out for pulse measurement in other locations. Fig.4 (d) shows the volunteer's wrist pulse signal measured at position A2 after exercise. The pulse rate increased from 90 beats per minute (at rest) to 138 beats per minute (after exercise), accompanied by a pronounced enhancement in signal amplitude, which is consistent with the anticipated physiological response.

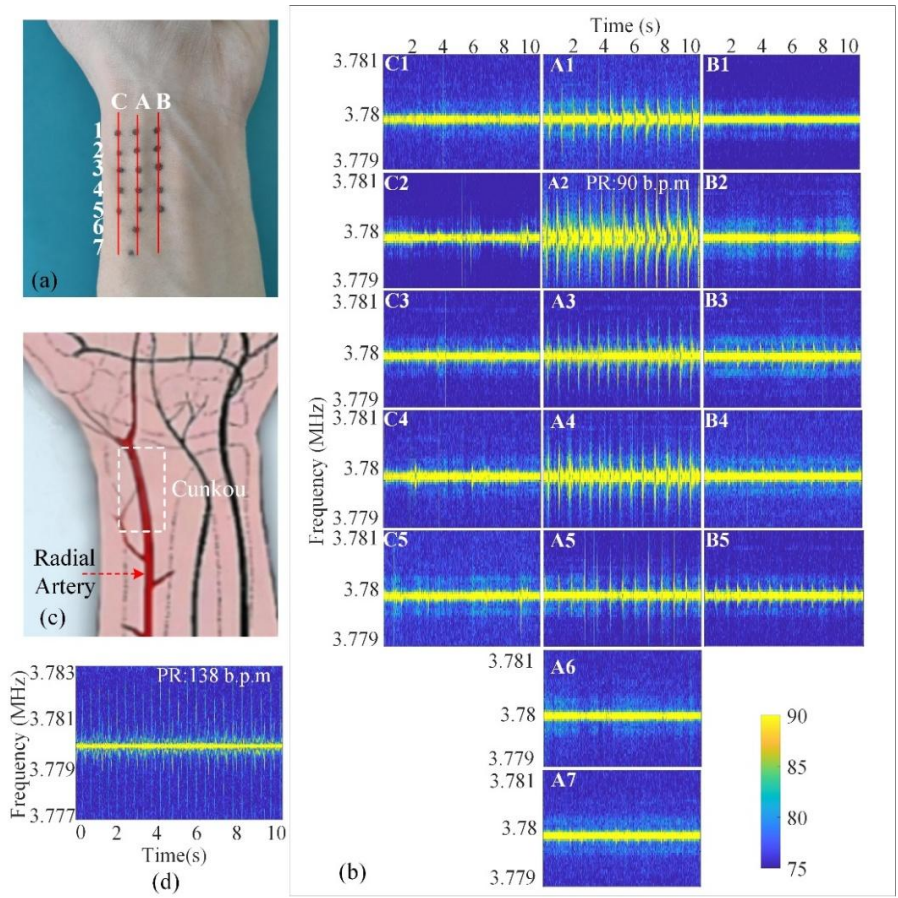

Fig. 4 Experimental results investigating the source of micro-motion signals at the wrist. (a) 17 test points around the wrist. (b) Micro-motion waveforms at different positions in resting. The measured pulse rate (PR) at position A2 in resting is 90 beats per minute (b.p.m). (c) Schematic illustrating the position of Cunkou near radial artery. (d) Pulse wave measurement results at position A2 after exercise and the measured PR is 138 b.p.m.

To validate the efficacy of the proposed system in detecting multiple human vital signs, the experimental setup was expanded with two parallel access points to obtain vital signs of two individuals simultaneously. Two young male volunteers were recruited to participate in this experiment. The actual environment was depicted in Fig. 5(a). The fiber lengths for the two pulse access points differed by 5 m. The two respiratory access points were situated within a microwave anechoic chamber, spaced approximately 0.5 meters apart without frequency-domain mutual obstruction. The acquired pulse and respiratory signals are presented in Fig. 5(b) and (c). The measured pulse rate for volunteer1 was 66 beats per minute with a respiratory rate of 14 breaths per minute and the volunteer2 exhibited a pulse rate of 64 beats per minute with a respiratory rate of 11 breaths per minute. It's worth noting that the measured pulse wave of the two volunteers revealed a frequency shift of 1.62 MHz, which is caused by the difference in the length of the fiber link between the two access points. To assess the accuracy of the proposed method, multiple reproducibility experiments were conducted. The pulse reference data were obtained using a commercially available smart bracelet (Dr.$O_2$ WS20A, Accurate.), with the pulse rate accuracy of ±3b.p.m. or 2%, whichever is greater. The respiratory reference data were derived from manual counts performed by the volunteers during the tests. The experiment was repeated 10 times under identical conditions, and the aggregated results are shown in Fig. 5(d). The test results demonstrated excellent agreement with the reference values, achieving an average accuracy of 98.1% for pulse measurements and 99.85% for respiratory measurements. The experimental results demonstrate that the proposed method can accurately measure multiple vital signs (respiratory rates and pulse rates) of multiple individuals simultaneously. The experiment also validated that the proposed system possesses the characteristic of distributed access. By reasonably configuring the low-loss optical fiber lengths connected to each collimator and in combination with the ubiquity of microwaves, a distributed effective multiple vital signs detection system has been proposed, as shown in Fig.1 (a). This system can distinguish respiratory and pulse signals from multiple individuals in the frequency domain, enabling real-time monitoring of multiple vital signs.

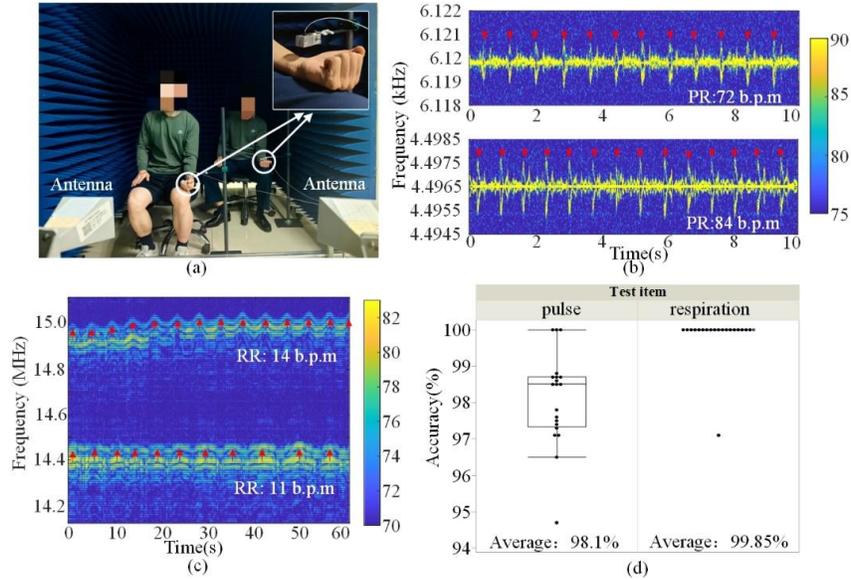

Fig. 5 Simultaneous measurement results of respiration and pulse of two individuals. (a)Multiple vital signs detection configuration. (b) Pulse measurement results. Each red triangle denotes a single pulse beat. PR: Pulse Rate; b.p.m: beats per minute. (c) Respiratory measurement results. Each red arrow indicates a single respiratory cycle. RR: Respiratory Rate; b.p.m: breaths per minute. (d) Accuracy of proposed system ( $Accuracy = 100\% \times |Measurement - \text{Reference}| / \text{Reference}$ ).

Early medical classic such as *The Yellow Emperor's Inner Canon* describes traditional Chinese pulse diagnosis as encompassing not only the radial artery but also the carotid and dorsalis pedis arteries, providing a more comprehensive assessment of health. Thus, our study extends beyond radial pulse measurements to include non-contact monitoring of the right carotid and right dorsalis pedis arteries, as shown in Fig.6 (a) and (b). This multi-site pulse signal acquisition approach, integrated with the theory of TCM, is anticipated to offer a novel research paradigm for TCM diagnosis. For non-contact detection, nonfixed monitoring positions help improve the universality of the monitoring system. Fig.6 (c) and (d) show the measured pulse signals at the superficial temporal artery and the right subclavian artery, which are commonly exposed in daily life. We performed pulse measurements at different arterials of one volunteer sequentially while the volunteer maintained a steady resting state, with each site measured for 10 seconds. The results show that the pulse rates measured at different arteries within the same time period were almost identical (about 11 to 13 beats within 10 seconds), indicating the universal applicability of the proposed method for pulse rate (or heart rate) monitoring at most major human arteries. Notably, the measured pulse signals at the right subclavian artery and the right dorsalis pedis artery were conspicuously weaker than those at other two arteries. The subclavian artery's deeper subcutaneous position results in greater signal attenuation before reaching the surface. Although superficially located, the dorsalis pedis

artery—as a distal lower limb vessel—experiences substantial pulse pressure decay due to prolonged vascular transmission and peripheral resistance. As a result, the surface-perceivable pulse amplitudes of the subclavian artery and the dorsalis pedis artery are significantly smaller than that of the carotid artery under the same physiological conditions [22]. Among all measured sites, the carotid artery exhibited the highest pulse amplitude and optimal signal-to-noise ratio, attributable to its superficial anatomical location, proximity to the heart, and larger vessel diameter. Fig.6 (e) presents high-resolution waveforms of three cardiac cycles from the carotid pulse signal. Each cycle displays distinct multi-peak characteristics, reflecting hemodynamic and cardiac properties. Detailed analysis of these waveforms may offer a novel approach for assessing physiological health status. The dots in Fig.6(f) illustrate the distribution of palpable arterial pulse points on the body surface, where green dots denote the sites measured in this work. The ability to measure pulse signals at different positions on the human body indicates that the proposed system can adapt to different body postures, as long as there is a bare leakage of pulse positions. Especially with the help of 3D mechanical positioning devices or optical phased arrays [26], the universality of the proposed method can be greatly expanded.

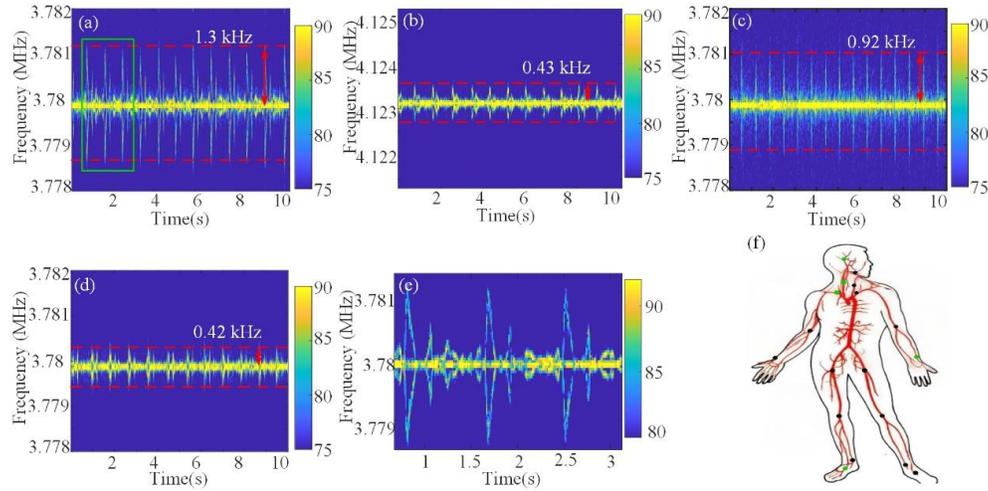

Fig.6 Pulse measurement at other arteries. (a) Pulse signal of the right carotid artery. (b) Pulse signal of the right dorsalis pedis artery. (c) Pulse signal of the superficial temporal artery. (d) Pulse signal of the right subclavian artery. (e) The enlarged view within the green box in (a). (f) The distribution of the main arteries in human body. The dots illustrate the distribution of palpable arterial pulse points on the body surface, where green dots denote the measurement sites investigated in this study.

As is well-established in physiology, both arterial pulse and blood pressure are induced by cardiac contraction and relaxation [22]. To investigate the correlation between measured pulse signals and blood pressure, we measured the pulse signals of the carotid artery under different blood pressure conditions. The measured pulse signal indicates two distinct peaks within each cycle, as shown in Fig. 7(a). The first peak corresponds to the maximum outward expansion velocity of the skin during systolic phase, while the secondary peak corresponds the maximum rebound velocity during diastolic phase. To induce varying blood pressure states, a volunteer performed exercises at different intensities. Immediately following each exercise, both blood pressure measurement and 10-second pulse signal acquisition were completed. The blood pressure was obtained by a clinically validated arm electronic sphygmomanometer (YE670A, Yuwell). The measured pulse signals of the volunteer with normal blood pressure (106/67 mmHg) and hypertensive blood pressure (133/75 mmHg) are presented in Fig. 7(a) and Fig. 7(b), respectively. After extracting the envelope of the pulse signal shown in Fig.7(a), the frequencies corresponding to the peaks of pulse signals can be clearly identified, as shown in Fig.7(c). The peak frequencies of each pulse from pulse signals across different blood pressure were extracted by this analytical approach. The average amplitude of the first peak within each cycle of the pulse signal were 0.687 MHz and 1.59 MHz over 10 seconds under 106/67 mmHg and 133/75 mmHg conditions, respectively. The pulse signal amplitude was significantly higher in hypertensive state compared to normotensive state. As indicated by Eq. (9), the amplitude of the pulse signal corresponds to the Doppler frequency of micro-vibrations. The larger amplitude of the pulse signal peaks suggests greater vibration, which in turn reflects increased blood flow velocity and potentially higher systolic blood pressure. Therefore, using the amplitude of the pulse signal peaks to characterize blood pressure levels has been proposed. We measured pulse signals under four different blood pressure conditions, with each acquisition lasting 10 seconds. Subsequent analysis established a quantitative relationship between the amplitude characteristics of pulse signal peaks and systolic blood pressure, as shown in Fig.7(d). Each data point in the figure represents the amplitude of the first peak in each pulse cycle. As blood pressure increased, the pulse rate rose, leading to a larger number of pulses within the same time window. Linear regression analysis performed using JMP software demonstrated that the amplitude of the first pulse peak showed a positive correlation with systolic pressure. The coefficient of determination ($R^2$) is 0.905, indicating the linear model provides a good fit to the experimental data. The proposed system enables effective blood pressure estimation based on the average amplitude of a 10-second pulse signal using the linear regression analysis. With further optimization of measurement stability, the system could potentially estimate blood pressure using the amplitude of a single pulse waveform, reducing the measurement duration to a single cardiac cycle. It will be a significant improvement compared to the conventional blood pressure measurement techniques with the measurement duration over ten seconds [27]. The proposed method may provide a promising solution for rapid blood pressure estimation and an effective reference for estimating the functional state of the cardiovascular system, which hold substantial clinical relevance for timely diagnosis of blood pressure related disorders, such as hypertension, severe hemorrhage.

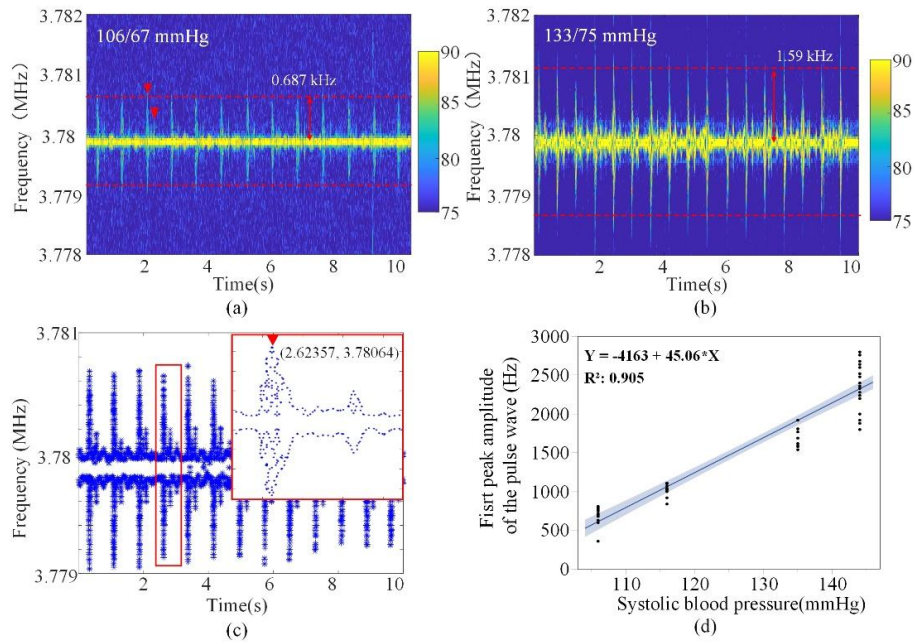

Fig.7 Relationship between the pulse wave and the blood pressure. (a) The pulse wave signal measured based on the proposed method with normal blood pressure (106/67 mmHg). The red triangles mark the two peaks of the pulse wave. (b) The measured pulse wave signal with hypertensive blood pressure (133/75 mmHg). (c) The blue line is the envelope of the measured pulse wave signal. Locally amplified signal is represented in the red box. (d) Relationship between the first peak amplitude of pulse wave and systolic blood pressure.

## 5. Discussion

Pulse diagnosis, as one of the cornerstone techniques in TCM, enables precise disease identification and treatment through the analysis of pulse waveform characteristics. Conventional TCM practice primarily relies on the cunkou diagnostic method, which partitions the radial artery region at the wrist into three distinct segments ("cun," "guan," and "chi"). These segments exhibit specific correspondences to different visceral states in bilateral distributions, with defined correlations between particular pulse patterns and pathological manifestations. Our study demonstrates that the proposed pulse acquisition system achieves synchronous multi-point measurements at the cunkou region. Through three independent sensing access point, the system accurately captures pulse signals from the chi, guan, and cun positions, thereby constructing comprehensive pulse patterns for clinical diagnostics. Compared with traditional palpation methods, this system offers two key advantages: (i) its non-contact measurement modality minimizes interference from patient anxiety on pulse waveforms, and (ii) it transforms subjective tactile perceptions into quantifiable objective data, significantly enhancing the accuracy of pulse characterization. This will significantly reduce the experience requirements for doctors in TCM palpation and lay a technical foundation for next-generation intelligent TCM diagnosis. Notably, we identify a statistically significant correlation between the pulse wave amplitude parameters obtained by our system and systolic blood pressure measurements. This finding confers dual diagnostic value to the platform: it not only facilitates TCM-based syndrome differentiation through pulse pattern analysis but also provides blood pressure evaluation in western medicine. Such dual functionality establishes a novel technological paradigm for integrative medicine approaches.

We present a distributed system capable of simultaneous multi-target vital sign monitoring, with theoretical modeling indicating simultaneous detection of hundreds of targets. We envisage applying such a high-performance, distributed system in a range of healthcare scenarios, such as round-the-clock vital sign monitoring in aged-care facilities, hospitals and custodial settings. Leveraging the cost-effectiveness of optical fibers, the system architecture enables large-scale implementation with centralized data processing and display terminals for integrated multi-target monitoring management. This architecture offers medical institutions an economically viable solution for comprehensive vital sign surveillance. In clinical practice, this system significantly enhances nursing efficiency by eliminating the need for traditional bedside measurements. Medical staff can access real-time vital sign data for any patient through centralized monitoring terminals, with the system automatically generating comprehensive monitoring logs. When integrated with intelligent early-warning algorithms, the system can promptly identify physiological abnormalities, thereby facilitating timely clinical intervention and potentially improving patient outcomes.

Beyond healthcare applications, we identify significant potential for this system in public security domains. Our findings demonstrate that pules signals can be reliably acquired from exposed aortic regions on the body surface without physical contact, which provides a technology for developing non-contact physiological monitoring systems in public settings. The non-contact characteristic preserves personal privacy and user comfort while providing public security and medical intelligence. The system permits the detection of potential threat actors through characteristic physiological stress signatures and the identification of individuals experiencing acute medical emergencies via abnormal vital parameters. When deployed in high-traffic public spaces, this system could enable a novel paradigm of proactive security screening and emergency medical response, potentially mitigating both intentional threats and spontaneous public health crises through early detection.

In summary，this work proposed a dual-domain MWP radar system for non-contact multiple vital signs detection, which represents a conceptual advance by extending non-contact vital sign monitoring into clinical diagnostics. Future studies should explore the system's generalizability across diverse patient populations and its potential integration with other digital health technologies.

## 6. Conclusion

This work proposed a dual-domain MWP radar for non-contact monitoring of multiple vital signs，in which the optical signals are used for pulse monitoring and the microwave signals are used for respiratory monitoring. Experimental results validate the distributed capability for simultaneous multiple vital signs monitoring of multiple individuals. The method can be used to monitor the pulse rate around the major arterial and the acquired pulse waveforms could reflect the blood pressure levels. It represents a conceptual advance by extending non-contact vital sign monitoring into clinical diagnostics. Future studies will explore the system's generalizability across diverse patient populations and its potential integration with other intelligent digital health technologies.

**Funding.** Natural Science Foundation of Hubei Province (ZRMS2023 002074); National Natural Science Foundation of China (62301602).

**Disclosures.** The authors declare no conflicts of interest.

**Data availability.** Data underlying the results presented in this paper are not publicly available at this time but may be obtained from the authors upon reasonable request.